\documentclass[traditabstract]{aa} 
\usepackage{graphicx}
\usepackage{txfonts}
\newcommand{\rc}{$^{12}$C/$^{13}$C}
\newcommand{\rn}{$^{14}$N/$^{15}$N}
\newcommand{\ro}{$^{16}$O/$^{18}$O}
\newcommand{\roh}{$^{16}$OH/$^{18}$OH}
\newcommand{\rdoh}{OD/OH}
\newcommand{\com}{C/2002 T7}

\begin{document}

\title{The \roh\ and \rdoh\ isotope ratios in comet C/2002 T7
(LINEAR)
\thanks{Based on observations collected at the
European Southern Observatory, Paranal, Chile (ESO Programme
073.C-0525).}}

\author{%
D. Hutsem\'ekers \inst{1}\fnmsep\thanks{DH is Senior Research Associate FNRS;
        JM is Research Director FNRS; and EJ is Research Associate FNRS}\and
J. Manfroid      \inst{1,\star\star}\and
E. Jehin         \inst{1,\star\star}\and
J.-M.~Zucconi    \inst{2}\and
C. Arpigny       \inst{1}}

\institute{%
Institut d'Astrophysique et de G\'eophysique,
Universit\'e de Li\`ege, All\'ee du 6 ao\^ut 17, B-4000 Li\`ege
\and
Observatoire de Besan\c{c}on, F25010 Besan\c{c}on Cedex, France
}

\date{}

\abstract{The \roh\ and \rdoh\ isotope ratios are measured in the
Oort-Cloud comet C/2002 T7 (LINEAR) through ground-based observations
of the OH $\, A\,^{2}\Sigma^{+} - X\,^{2}\Pi_{i}$ ultraviolet bands at
3063~\AA\ (0,0) and 3121~\AA\ (1,1) secured with the Very Large
Telescope (VLT) feeding the Ultraviolet-Visual Echelle Spectrograph
(UVES).  From the \roh\ ratio, we find \ro\ = 425 $\pm$ 55, equal
within the uncertainties to the terrestrial value and to the ratio
measured in other comets, although marginally smaller.  We also
estimate \rdoh\ from which we derive D/H = 2.5 $\pm$ 0.7 10$^{-4}$ in
water. This value is compatible with the water D/H ratios evaluated in
other comets and marginally higher than the terrestrial value.
}


\maketitle
%

\section{Introduction}
\label{sect:intro}

The determination of the abundance ratios of the stable isotopes of
the light elements in different objects of the Solar System provides
important clues into the study of their origin and history.  This is
especially true for comets which carry the most valuable information
regarding the material in the primitive solar nebula.

The \ro\ isotopic ratio has been measured from space missions in a few
comets. In-situ measurements with the neutral and ion mass
spectrometers onboard the Giotto spacecraft gave \ro\ = 495$\pm$37 for
H$_2$O in comet 1P/Halley (Eberhardt et al. \cite{Eberhardt}). A deep
integration of the spectrum of the bright comet 153P/2002 C1
(Ikeya-Zhang) with the sub-millimeter satellite Odin led to the
detection of the H$_2^{18}$O line at 548 GHz (Lecacheux et
al. \cite{Lecacheux}).  Subsequent observations resulted in the
determination of \ro\ = 530$\pm$60, 530$\pm$60, 550$\pm$75 and
508$\pm$33 in the Oort-Cloud comets Ikeya-Zhang, C/2001 Q4, C/2002 T7
and C/2004 Q2 respectively (Biver et al. \cite{Biver}).  Within the
error bars, these measurements are consistent with the terrestrial
value (\ro\ {\small(SMOW\footnote{Standard Mean Ocean Water})} = 499),
although marginally higher (Biver et al. \cite{Biver}).  More
recently, laboratory analyses of the silicate and oxide mineral grains
from the Jupiter family comet 81P/Wild~2 returned by the Stardust
space mission provided \ro\ ratios also in excellent agreement with
the terrestrial value. Only one refractory grain appeared marginally
depleted in $^{18}$O (\ro\ = 576$\pm$78) as observed in refractory
inclusions in meteorites (McKeegan et al.~\cite{McKeegan}).

The D/H ratio has been measured in four comets. In-situ measurements
provided D/H = 3.16$\pm$0.34 10$^{-4}$ for H$_2$O in 1P/Halley
(Eberhardt et al. \cite{Eberhardt}, Balsiger et al. \cite{Balsiger}),
a factor of two higher than the terrestrial value (D/H {\small (SMOW)}
= 1.556 10$^{-4}$).  The advent of powerful sub-millimeter telescopes,
namely the Caltech Submillimeter Observatory and the James Clerck
Maxwell telescope located in Hawaii, allowed the determination of the
D/H ratio for two exceptionally bright comets.  In comet C/1996 B2
(Hyakutake), D/H was found equal to 2.9$\pm$1.0 10$^{-4}$ in H$_2$O
(Bockel\'ee-Morvan et al. \cite{Bockelee}), while, in comet C/1995 O1
(Hale-Bopp), the ratios D/H = 3.3$\pm$0.8 10$^{-4}$ in H$_2$O and D/H
= 2.3$\pm$0.4 10$^{-3}$ in HCN were measured (Meier et
al. \cite{Meier1,Meier2}), confirming the high D/H value in
comets. Both Hyakutake and Hale-Bopp are Oort-Cloud comets.  Finally,
bulk fragments of 81P/Wild~2 grains returned by Stardust indicated
moderate D/H enhancements with respect to the terrestrial
value. Although D/H in 81P/Wild~2 cannot be ascribed to water, the
measured values overlap the range of water D/H ratios determined in
the other comets (McKeegan et al.~\cite{McKeegan}).

Among a series of spectra obtained with UVES at the VLT to measure the
\rn\ and \rc\ isotope ratios in various comets from the 3880$\,$\AA\
CN ultraviolet band (e.g. Arpigny et al. \cite{Arpigny}, Hutsem\'ekers
et al. \cite{Hutsemekers}, Jehin et al. \cite{Jehin2}, Manfroid et
al. \cite{Manfroid2}), we found that the spectrum of \com\ appeared
bright enough to detect the $^{18}$OH lines in the $A\,^{2}\Sigma^{+}
- X\,^{2}\Pi_{i}$ bands at 3100~\AA\ allowing --for the first time--
the determination of the \ro\ ratio from {\it ground-based}
observations.  We also realized that the signal-to-noise ratio of our
data was sufficient to allow a reasonable estimate of the \rdoh\ ratio
from the same bands.

The possibility of determining the \ro\ ratio from the OH ultraviolet
bands has been emphasized by Kim (\cite{Kim}). Measurements of the
OD/OH ratio were already attempted by A'Hearn et al. (\cite{AHearn})
using high resolution spectra from the International Ultraviolet
Explorer and resulting in the upper limit D/H $< 4 \, 10^{-4}$ for
comet C/1989 C1 (Austin). These observations now become feasible from
the ground thanks to the high ultraviolet throughput of spectrographs
like UVES at the VLT.

\section{Observations and data analysis}            

Observations of comet \com\ were carried out with UVES mounted on the
8.2m UT2 telescope of the European Southern Observatory VLT.  Spectra
in the wavelength range 3040$\,$\AA --10420$\,$\AA\ were secured in
service mode during the period May~6, 2004 to June~12, 2004. The UVES
settings 346+580 and 437+860 were used with dichroic \#1 and \#2
respectively. In the following, only the brighest ultraviolet spectra
obtained on May~6, May~26 and May~28 are considered.  The 0.44
$\times$ 10.0 arcsec slit provided a resolving power $R \simeq 80000$.
The slit was oriented along the tail, centered on the nucleus on
May~26, and off-set from the nucleus for the May~6 and May~28
observations. The observing circumstances are summarized in
Table~\ref{tab:obs}.
 
\begin{table}[t]
\caption{Observing circumstances}
\label{tab:obs}
\begin{tabular}{lcccccr}\hline\hline \\[-0.10in]
Date   & $r$  & $\dot{r}$     & $\Delta$ & Offset  & $t$ & Airmass \\
(2004) & (AU) & (km/s) & (AU) & (10$^{3}$ km) & (s)        &  \\
       \hline \\[-0.10in]
May 6  & 0.68  & 15.8 & 0.61 & 1.3 & 1080  & 2.2-1.9   \\
May 26 & 0.94  & 25.6 & 0.41 & 0.0 & 2677  & 1.3-1.8   \\
May 26 & 0.94  & 25.6 & 0.41 & 0.0 & 1800  & 2.1-2.7   \\
May 28 & 0.97  & 25.9 & 0.48 & 10.0 &3600  & 1.3-1.7   \\
\hline\\[-0.2cm]
\end{tabular}\\
{\footnotesize $r$ and $\dot{r}$ are the comet heliocentric distance
and radial velocity; $\Delta$ is the geocentric distance; $t$ is the
exposure time; Airmass is given at the beginning and at the end of the
exposure}
\end{table}

The spectra were reduced using the UVES pipeline (Ballester et
al. \cite{Ballester}), modified to accurately merge the orders taking
into account the two-dimensional nature of the spectra. The
flat-fields were obtained with the deuterium lamp which is more
powerful in the ultraviolet.

The data analysis and the isotopic ratio measurements were performed
using the method designed to estimate the carbon and nitrogen isotopic
ratios from the CN ultraviolet spectrum (Arpigny et
al. \cite{Arpigny}, Jehin et al. \cite{Jehin} and Manfroid et
al. \cite{Manfroid}).  Basically, we compute synthetic fluorescence
spectra of the $^{16}$OH, $^{18}$OH and $^{16}$OD for the
$A\,^{2}\Sigma^{+} - X\,^{2}\Pi_{i}$ (0,0) and (1,1) ultraviolet bands
for each observing circumstance.  Isotope ratios are then estimated by
fitting the observed OH spectra with a linear combination of the
synthetic spectra of the two species of interest.

\subsection{The OH model}

We have developed a fluorescence model for OH similar to the one
described by Schleicher and A'Hearn (\cite{Schleicher}). As lines of
the OH(2-2) bands are clearly visible in our spectra we have included
vibrational states up to $v=2$ in the A$^2\Sigma^+$ and X$^2\Pi_i$
electronic states. For each vibrational state rotational levels up to
$J=11/2$ were included, leading to more than 900 electronic and
vibration-rotation transitions. The system was then solved as
described in Zucconi and Festou (\cite{Zucconi}).

Accurate OH wavelengths were computed using the spectroscopic
constants of Colin et al. (\cite{Colin}) and Stark et
al. (\cite{Stark}). OD wavelengths were computed using the
spectroscopic constants of Abrams et al. (\cite{Abrams}) and Stark et
al. (\cite{Stark}). $^{18}$OH wavelengths were derived from the
$^{16}$OH ones using the standard isotopic shift formula; they are
consistent with the measured values of Cheung et al. (\cite{Cheung}).

Electronic transition probabilities for OH and OD are given by Luque
and Crosley (\cite{Luque1,Luque2}). We used the dipole moments of OH
and OD measured by Peterson et al. (\cite{Peterson}) to compute the
rotational transition probabilities and the vibrational lifetimes
computed by Mies (\cite{Mies}). Because of the very small difference
in the structure of $^{18}$OH and $^{16}$OH the transition
probabilities for $^{18}$OH and $^{16}$OH are the same.

The OH fluorescence spectrum is strongly affected by the solar
Fraunhofer lines, especially in the 0-0 band, so a carefully
calibrated solar atlas is required. We have used the Kurucz
(\cite{Kurucz}) atlas above 2990~\AA\ and the A'Hearn et
al. (\cite{AHearn1}) atlas below.

 The role of collisions in the OH emission, in particular those with
charged particles inducing transitions in the $\Lambda$ doublet ground
rotational state, was first pointed out by Despois et
al. (\cite{Despois}) in the context of the 18~cm radio emission and
then also considered in the UV emission by Schleicher
(\cite{SchleicherPHD}, Schleicher and A'Hearn \cite{Schleicher}).
Modeling the effect of collisions may be done by adding the collision
probability transition rate between any two levels, $i$ and $j$:
$$C_{i,j} = \sum_c{n_c({\bf r})\,{\rm v}_c({\bf r})\,\sigma_c(i,j,{\rm v}_c)}$$
where the sum extends over all colliders. $n_c$ is the local density
of the particles inducing the transition, ${\rm v}_c$ is the relative
velocity of the particles and $\sigma_c$ is the collision cross
section. It also depends on the energy of the collision i.e. of ${\rm
v}_c$.  The reciprocal transition rates are obtained through detailed
balance:
$$C_{j,i} = C_{i,j}\frac{g_i}{g_j}\exp(E_{ij}/kT)$$
in which $g_i$ is the statistical weight and $E_{ij}$ is the energy
separation between the states.  In order to reduce the number of
parameters required to model the collisions we have adopted a
simplified expression of the form $C_{i,j} = q_\Lambda$ for the
transition in the $\Lambda$ doublet ground state.  In order to better
fit the OH spectra we have also found necessary to take into account
rotational excitation through a similar expression $C_{i,j} = q_{rot}$
with $q_{rot}$ different from 0 only for dipole transitions, i.e. when
$\Delta J < 2$, which appeared to correctly fit the data. Furthermore,
since OH and OD have similar dipole moments, we assumed that
collisional cross-sections are identical for both molecules.

The model assumes that the $^{16}$OH lines are optically thin.  This
is verified by the fact that it correctly reproduces both the faint
and strong OH emission lines.

\subsection{\roh}

Two $^{18}$OH lines at 3086.272 \AA\ and 3091.046 \AA\ are clearly
detected in the (0,0) band. However these lines are strongly blended
with the $\sim$ 500 times brighter $^{16}$OH emission lines and then
not useful for an accurate flux estimate.  In fact the (1,1) band at
3121~\AA{}, while fainter, is better suited for the determination of
\roh\ since (i) the wavelength separation between $^{18}$OH and
$^{16}$OH is larger ($\simeq$ 0.3 \AA\ instead of 0.1 \AA{}), and (ii)
the sensitivity of UVES rapidly increases towards longer wavelengths
while the atmospheric extinction decreases, resulting in a better
signal-to-noise ratio.

\begin{figure}[t]
\resizebox{\hsize}{!}{\includegraphics*{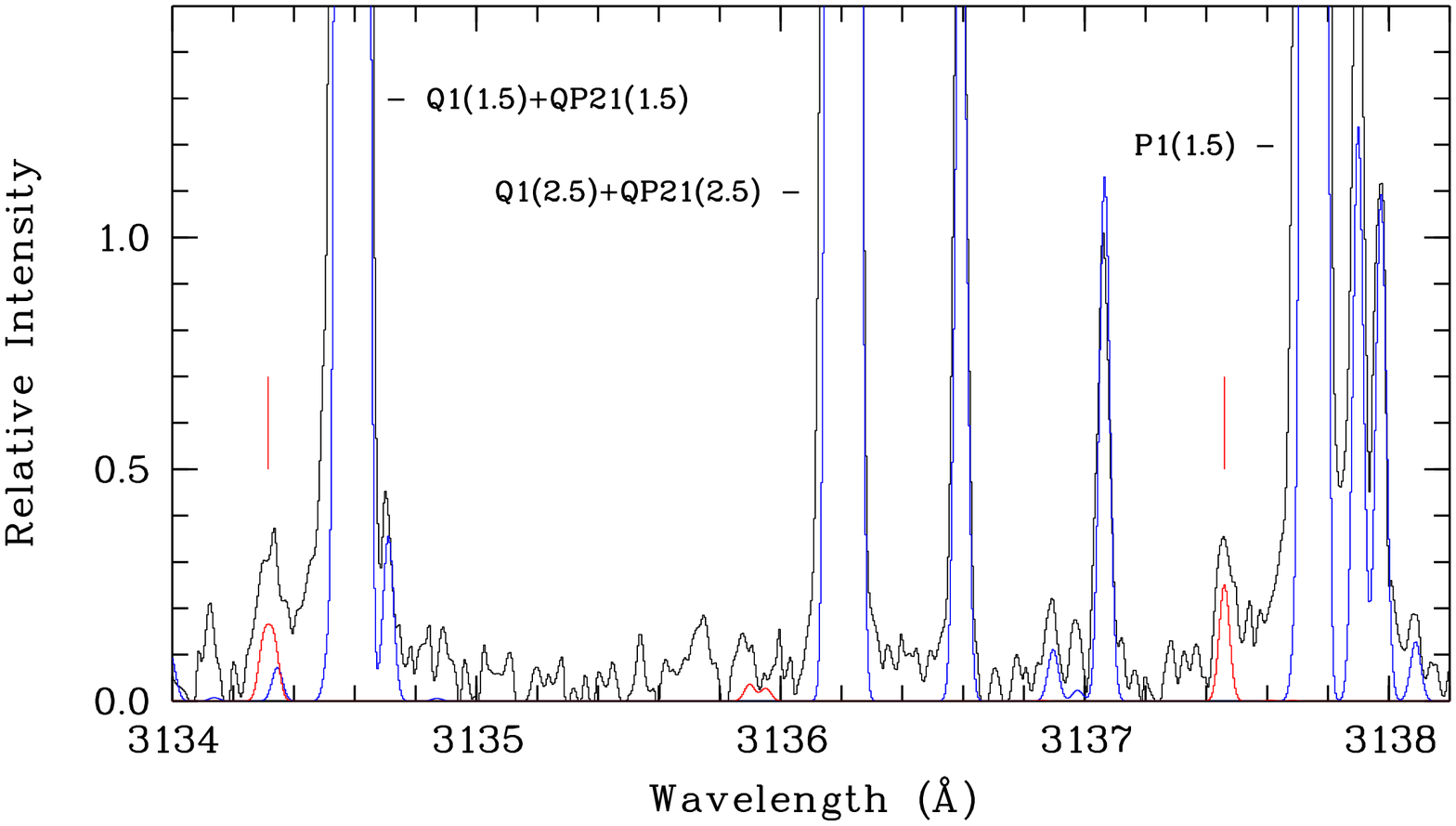}}
\caption{A section of the May 6 spectrum of the OH (1,1) band in comet
\com .  {\it Black:} observed spectrum; {\it blue:} synthetic emission
spectrum of $^{16}$OH fitted to the observed one (the extended wings
of the $^{16}$OH lines are not modelled here; on the other hand, in
this and all three other figures, the presence of prompt emission is
taken into account); {\it red:} synthetic fluorescence spectrum of
$^{18}$OH with the typical ratio \roh\ = 500. The position of the
$^{18}$OH lines at 3134.315 \AA\ and 3137.459 \AA\ is indicated.}
\label{fig:fig1}
\end{figure}

Fig.~\ref{fig:fig1} illustrates a part of the observed OH (1,1) band
together with the synthetic spectrum from the model.  Two $^{18}$OH
lines are clearly identified.

To actually evaluate \roh\ we first select the 3 brighest and best
separated $^{18}$OH lines at $\lambda$ = 3134.315$\,$\AA ,
3137.459$\,$\AA\ and 3142.203$\,$\AA .  These lines are then
doppler-shifted and co-added with proper weights to produce an average
profile which is compared to the $^{16}$OH profile similarly treated
(cf. Jehin et al. \cite{Jehin} for more details on the method).  We
verified that the $^{16}$OH faint wings and nearby prompt emission
lines (analysed in detail in a forthcoming paper) do not contaminate
the $^{18}$OH lines nor the measurement of the isotopic ratios.  The
ratio \roh\ is then derived through an iterative procedure which is
repeated for each spectrum independently.  For the spectra of May 6,
26 and 28 we respectively derive \roh\ = 410$\pm$60, 510$\pm$130 and
380$\pm$290.  The uncertainties are estimated from the co-added
spectra by considering the rms noise in spectral regions adjacent to
the $^{18}$OH lines, and by evaluating errors in the positioning of
the underlying pseudo-continuum (i.e. the dust continuum plus the
faint wings of the strong lines).  The weighted average of all
measurements gives \roh\ = 425 $\pm$ 55.

Since OH is essentially produced from the dissociation of H$_2$O,
\roh\ represents the \ro\ ratio in cometary water, with the reasonable
assumption that photodissociation cross-sections are identical for
H$_2$$^{18}$O and H$_2$$^{16}$O.

\begin{figure}[t]
\resizebox{\hsize}{!}{\includegraphics*{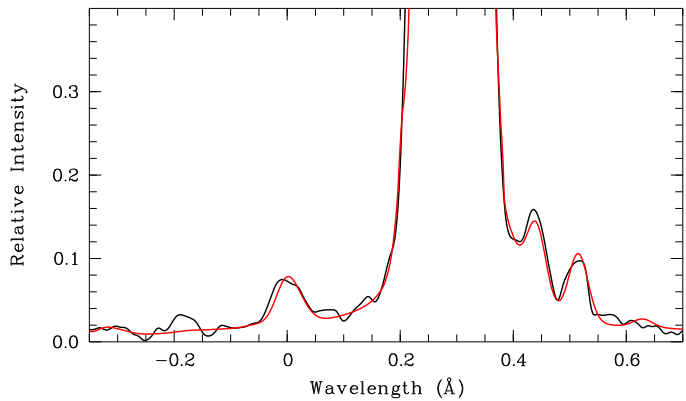}}
\caption{Co-addition of the $^{18}$OH lines at $\lambda$ =
3134.315$\,$\AA , 3137.459$\,$\AA\ and 3142.203$\,$\AA\ from the May 6
spectrum after proper wavelength shifts. Co-added $^{18}$OH is shifted
to $\lambda$=0 while $^{16}$OH appears at $\lambda$$ \simeq$ 0.3. {\it
Black:} observed spectrum; {\it red:} synthetic combined spectrum of
$^{16}$OH, $^{18}$OH with \roh\ = 410, and an empirical fit to the OH
extended wings.}
\label{fig:fig2}
\end{figure}

\subsection{\rdoh}

\begin{figure}[t]
\resizebox{\hsize}{!}{\includegraphics*{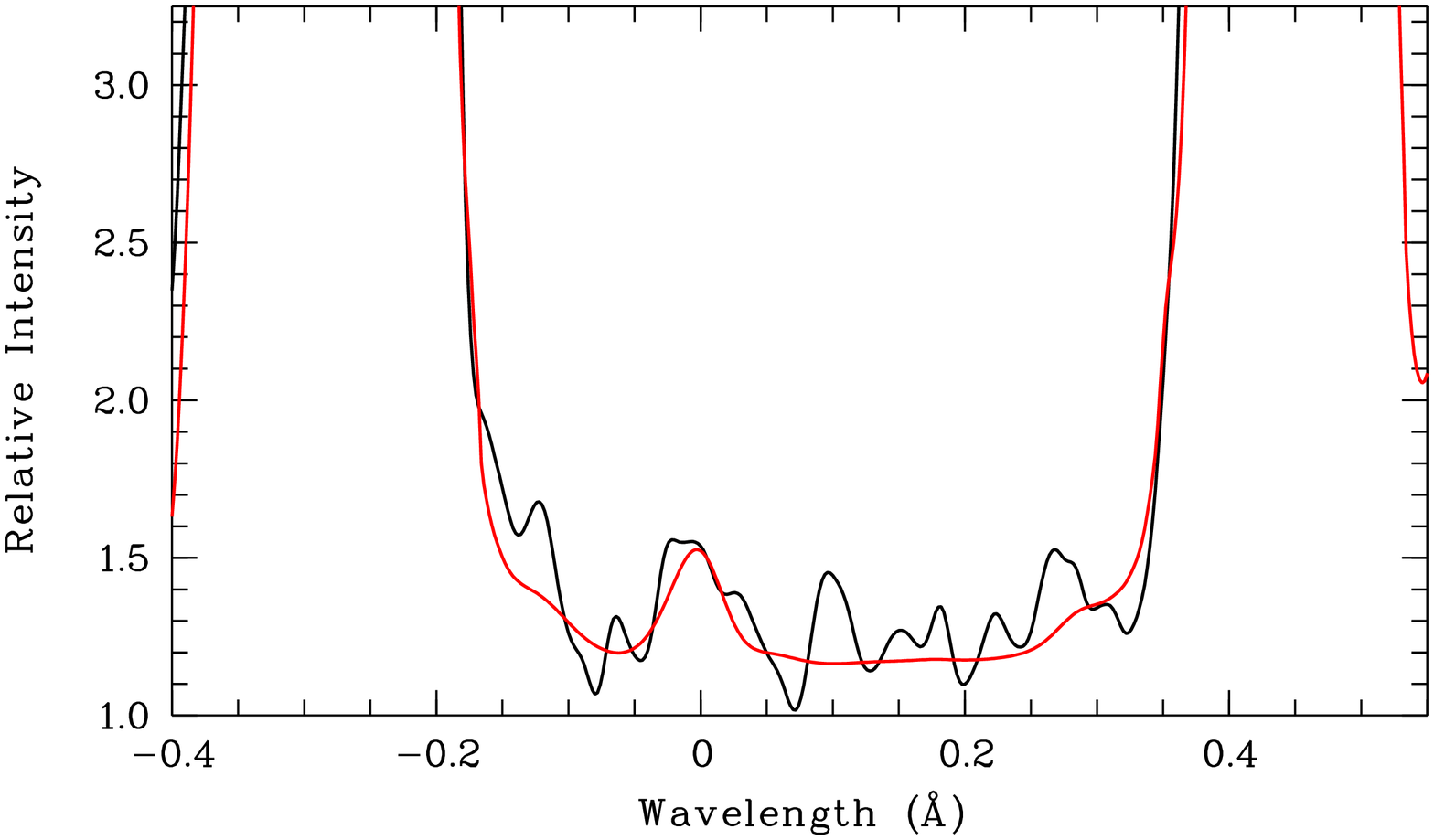}}
\caption{Co-addition of 27 OD lines from the May 6 spectrum after
proper wavelength shifts.  Co-added OD is shifted to
$\lambda$=0. {\it Black:} observed spectrum; {\it red:}
synthetic combined spectrum of OH, OD with \rdoh\ = 4 10$^{-4}$, and an
empirical fit to the OH extended wings.}
\label{fig:fig3}
\end{figure}

\begin{figure}[t]
\resizebox{\hsize}{!}{\includegraphics*{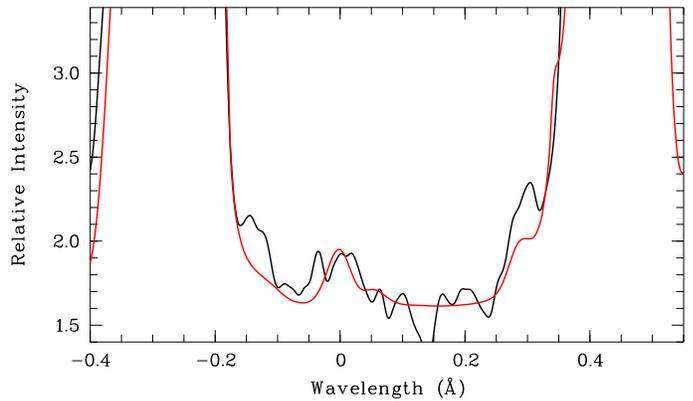}}
\caption{Same as Fig.~\ref{fig:fig3} for the May 26 spectrum.}
\label{fig:fig4}
\end{figure}

The detection of OD lines is much more challenging since one may
expect the OD lines to be a few thousand times fainter than the OH
lines.  Fortunately, the wavelength separation between OD and OH
($\gtrsim$ 10 \AA{}) is much larger than between $^{18}$OH and
$^{16}$OH such that both the (0,0) and (1,1) bands can be used with no
OD/OH blending (apart from chance coincidences).  Since no individual
OD lines could be detected, we consider the 30 brighest OD lines (as
predicted by the model) for co-addition.  After removing 3 of them,
blended with other emission lines, an average profile is built with
careful Doppler-shifting and weighting as done for $^{18}$OH.  Only
our best spectra obtained on May 6 and May 26 are considered, noting
that the (0,0) band --which dominates the co-addition-- is best
exposed on May 26 while the (1,1) band is best exposed on May 6, due
to the difference in airmass. The resulting OD line profiles are
illustrated in Fig.~\ref{fig:fig3} and \ref{fig:fig4} and compared to
a synthetic spectrum computed with \rdoh\ = 4 10$^{-4}$.  OD is
detected as a faint emission feature which is present {\it at both
epochs}. From the measurement of the line intensities, we derive
\rdoh\ = 3.3$\pm$1.1 10$^{-4}$ and 4.1$\pm$2.0 10$^{-4}$ for the
spectra obtained on May 6 and 26 respectively. The weighted average is
\rdoh\ = 3.5$\pm$1.0 10$^{-4}$.  The difference in the lifetime of OD
and OH (van Dishoeck and Dalgarno \cite{Vandishoeck}) does not
significantly affect our results since the part of the coma sampled by
the UVES slit is two orders of magnitude smaller than the typical OH
scale-length.  The uncertainties on \rdoh\ were estimated as for \roh
.  Possible errors on the isotopic ratios related to uncertainties on
the collision coefficients were estimated via simulations and found to
be negligible. Even in the hypothetical case that collisions
differently affect OD and OH, errors are much smaller than the other
uncertainties, as expected since the contribution of collisions is
small with respect to the contribution due to pure fluorescence.

To estimate the cometary D/H ratio in water, HDO/H$_2$O must be
evaluated.  While the cross-section for photodissociation of HDO is
similar to that of H$_2$O, the production of OD+H is favoured over
OH+D by a factor around 2.5 (Zhang and Imre \cite{Zhang}, Engel and
Schinke \cite{Engel}).  Assuming that the total branching ratio for
HDO $\rightarrow$ OD + H plus HDO $\rightarrow$ OH + D is equal to
that of H$_2$O $\rightarrow$ OH + H, we find HDO/H$_2$O $\simeq$ 1.4
OD/OH.  With D/H = 0.5 HDO/H$_2$O, we finally derive D/H = 2.5$\pm$0.7
10$^{-4}$ in cometary water.  The factor (OD+H)/(OH+D) = 2.5 adopted
in computing the branching ratios for the photodissociation of HDO is
an average value over the spectral region where the cross-sections
peak. In fact (OD+H)/(OH+D) depends on the wavelength and roughly
ranges between 2 and 3 over the spectral regions where absorption is
significant (Engel and Schinke \cite{Engel}, Zhang et
al. \cite{Zhang2}, Yi et al. \cite{Yi}).  Fortunately, even if we
adopt the extreme ratios (OD+H)/(OH+D) = 2 or (OD+H)/(OH+D) = 3
instead of 2.5, the value of the D/H isotopic ratio is not changed by
more than 6\%.

\section{Discussion} 

We have measured the oxygen isotopic ratio \ro\ = 425 $\pm$ 55 from
the OH \ $A\,^{2}\Sigma^{+} - X\,^{2}\Pi_{i}$ ultraviolet bands in
comet \com .  Although marginally smaller, our value do agree within
the uncertainties with \ro\ = 550 $\pm$ 75 estimated from observations
by the Odin satellite (Biver et al. \cite{Biver}), with the \ro\
ratios determined in other comets, and with the terrestrial value
(Sect.~\ref{sect:intro}).

To explain the so-called ``oxygen anomaly'' i.e. the fact that oxygen
isotope variations in meteorites cannot be explained by mass-dependent
fractionation, models of the pre-solar nebula based on CO
self-shielding were proposed, predicting enrichments, with respect to
the {\small SMOW} value, of $^{18}$O in cometary water up to \ro\
$\sim$ 415 (Yurimoto \& Kuramoto \cite{Yurimoto}).  Recently, Sakamoto
et al. (\cite{Sakamoto}) found evidence for such an enrichment in a
primitive carbonaceous chondrite, supporting self-shielding
models. The value of \ro\ we found in \com\ is also marginally smaller
than the terrestrial value and compatible with these predictions.  On
the other hand, the measurement of \ro\ = 440 $\pm$ 6 in the solar
photosphere (Ayres et al. \cite{Ayres}; cf. Wiens et al. \cite{Wiens}
for a review of other, less accurate, measurements) indicates that
solar ratios may deviate from the terrestrial ratios by much larger
factors than anticipated, requiring some revision of the models.  More
observations are then critically needed to get an accurate value of
\ro\ in comets, assuming that cometary water is pristine enough and
can be characterized by a small set of representative values. Namely,
if self-shielding is important in the formation of the solar system,
it is not excluded that significant variations can be observed between
comets formed at different locations in the solar system, like Oort
cloud and Jupiter-family comets.

We also detected OD and estimated D/H = 2.5 $\pm$ 0.7 10$^{-4}$ in
water. Our measurement is compatible with other values of D/H in
cometary water and marginally higher than the terrestrial value
(Sect.~\ref{sect:intro}).  Our observations were not optimized for the
measurement of OD/OH (neither for \roh ) and one of our best spectra
was obtained at airmass $\sim$ 2 with less than 20 min of exposure
time for a comet of heliocentric magnitude m$_r \simeq$ 5 (for
comparison, comet Hale-Bopp reached m$_r \simeq$ $-$1). All these
observing circumstances can be improved, including observations at
negative heliocentric velocities to increase the OD/OH fluorescence
efficiency ratio (cf. figure~1 of A'Hearn et al. \cite{AHearn}). This
opens the possibility to routinely measure both the \ro\ and D/H
ratios from the ground, together with the \rc\ and \rn\ ratios, for a
statistically significant sample of comets of different types
(e.g. Oort-cloud, Halley-type, and hopefully Jupiter-family comets
although the latter are usually fainter). The measurement of D/H is
especially important since it allows to limit the contribution of
comets to the terrestrial water, the high D abundance implying that no
more than about 10 to 30\% of Earth's water can be attributed to
comets (e.g. Eberhardt et al. \cite{Eberhardt}, Dauphas et
al. \cite{Dauphas}, Morbidelli et al. \cite{Morbidelli}).  However,
only a full census of D/H in comets could answer this question. In
particular, if Jupiter-family comets, thought to have formed in
farther and colder places in the Solar System, are characterized by an
even higher D/H, closer to the ratio measured in the interstellar
medium water, then the fraction of cometary H$_2$O brought onto the
Earth could be even smaller.

\begin{acknowledgements}
We thank the referee, Dominique Bockel\'ee-Morvan, for comments which
helped to significantly improve the manuscript.  We are also grateful
to Paul Feldman and Hal Weaver for useful discussions.
\end{acknowledgements}

\end{document}